\documentclass[%
 aip,
 amsmath,amssymb,
 reprint,%
]{revtex4-2}

\usepackage{graphicx}
\usepackage{dcolumn}
\usepackage{bm}

\usepackage[utf8]{inputenc}
\usepackage[T1]{fontenc}
\usepackage{mathptmx}
\usepackage{etoolbox}
\usepackage{color}
\makeatletter
\def\@email#1#2{%
 \endgroup
 \patchcmd{\titleblock@produce}
  {\frontmatter@RRAPformat}
  {\frontmatter@RRAPformat{\produce@RRAP{*#1\href{mailto:#2}{#2}}}\frontmatter@RRAPformat}
  {}{}
}%
\makeatother
\begin{document}

\preprint{AIP/123-QED}

\title{High thermoelectric performance in metastable phase of silicon: a first$-$principles study}
\author{Yongchao Rao, C. Y. Zhao, and Shenghong Ju$^{*}$}
\affiliation{China-UK Low Carbon College, Shanghai Jiao Tong University, Shanghai 201306, P. R. China.}
\email{shenghong.ju@sjtu.edu.cn}
\date{\today}

\begin{abstract}
In this work, both thermal and electrical transport properties of diamond$-$cubic Si (Si$-$I) and metastable R8 phase of Si (Si$-$XII) are comparatively studied by using first$-$principles calculations combined with Boltzmann transport theory. The metastable Si$-$XII shows one magnitude lower lattice thermal conductivity than stable Si$-$I from 300 to 500~K, attributed from the stronger phonon scattering in three$-$phonon scattering processes of Si$-$XII. For the electronic transport properties, although Si$-$XII with smaller band gap (0.22~eV) shows lower Seebeck coefficient, the electrical conductivities of anisotropic $n$$-$type Si$-$XII show considerable values along $x$ axis due to the small effective masses of electron along this direction. The peaks of thermoelectric figure of merit ($ZT$) in $n$$-$type Si$-$XII are higher than that of $p$$-$type ones along the same direction. Owing to the lower lattice thermal conductivity and optimistic electrical conductivity, Si$-$XII exhibits larger optimal $ZT$ compared with Si$-$I in both $p$$-$ and $n$$-$type doping. For $n$$-$type Si$-$XII, the optimal $ZT$ values at 300, 400, and 500~K can reach 0.24, 0.43, and 0.63 along $x$ axis at carrier concentration of $2.6\times10^{19}$, $4.1\times10^{19}$, and $4.8\times10^{19}$~cm$^{-3}$, respectively. The reported results elucidate that the metastable Si could be integrated to the thermoelectric power generator.
\end{abstract}

\maketitle
The design and synthesis of novel materials with considerable thermoelectric energy conversion efficiency have made great achievements during the past few decades \cite{1}. However, the conventional thermoelectric materials are usually expensive and contain rare or toxic elements, and most of them are always unstable at high temperature \cite{2}. The earth$-$abundant Si has been the fundamental material in nano$-$electronics industry, while the diamond$-$cubic Si suffers from extremely low thermoelectric figure of merit ($ZT$$\sim$0.01 at 500~K) due to its high bulk thermal conductivity ($k$$\sim$150~Wm${^{-1}}$K${^{-1}}$) \cite{3}. Aiming to modulate the transport of heat and charge carriers in thermoelectric materials, nano$-$engineering approaches including nanomeshes \cite{4}, nanoinclusions \cite{5}, nanowires \cite{6,7} and nanocrystals \cite{8} have been employed to suppress the thermal transportation, thus improving the $ZT$. However, the experimental synthesis of high density and highly uniform nanostructure is a key challenge for realizing high$-$performance silicon$-$based thermoelectric modules. For example, the weak mechanical strength and strong dependence on both diameter \cite{6} and surface morphology \cite{9,10} of Si nanowire greatly hinder its large$-$scale application. Therefore, studying Si$-$based materials with intrinsic low lattice thermal conductivity is of curial meaningful and necessary.

The pressure$-$induced metastable phases of Si attracted intense research interest and have been investigated in indentation experiments with precisely controlled strain rates \cite{11,12,13}. Under pressure around 12~GPa, the diamond$-$cubic structure of DC$-$Si (Si$-$I, $Fd\overline{3}m$) firstly transforms to metallic $\beta$$-$Sn phase (Si$-$II, $I4_{\rm 1}/amd$) at room temperature \cite{14}. Decompression from Si$-$II follows a different structural sequence depending on the pressure release rate. Slow decompression leads to semi$-$metallic R8 phase (Si$-$XII, $R\overline{3}$) at approximately 9.3~GPa and further to a metastable body$-$centered BC8 phase (Si$-$III, $Ia\overline{3}$) when pressure is completely released \cite{15}. Both Si$-$III and Si$-$XII phases persist at ambient pressure, which makes them possible to be synthesized. Compared with stable phase Si$-$I, the metastable Si phases show quite different electrical \cite{16}, optical \cite{17} and thermal \cite{18, 19} properties, which have potential applications in energy generation and storage, including the thermoelectrics, solar cells, batteries and catalysts. Chon et al. \cite{16} investigated the electrical properties of bulk nanograined Si$-$III and Si$-$XII fabricated by high$-$pressure torsion processing, and found that the resistivity can be controlled not only by adding dopants but also by the formation of metastable phases. By performing density functional and many body perturbation theory calculation, Wippermann et al.\cite{17} reported that Si$-$XII nanoparticles exhibited significantly lower electronic gaps and a redshifted optical absorption compared with Si$-$I, verifying Si$-$XII was a promising candidate for solar energy conversion with multiple exciton generation.

All of these researches motivate us to further explore the thermoelectric performance of Si$-$XII, and also there are still remaining open questions
to answer whether the thermoelectric performance of metastable Si is better than the stable one, and how it works. In this study, we conducted extensive investigation on phonon and electron transport properties and thermoelectric performance of both Si$-$XII and Si$-$I via the density-functional theory (DFT) calculations, aiming to address the unclear answer and elucidate the underlying physical mechanisms.

\begin{figure}[h]
  \centering
  \includegraphics[width=8cm]{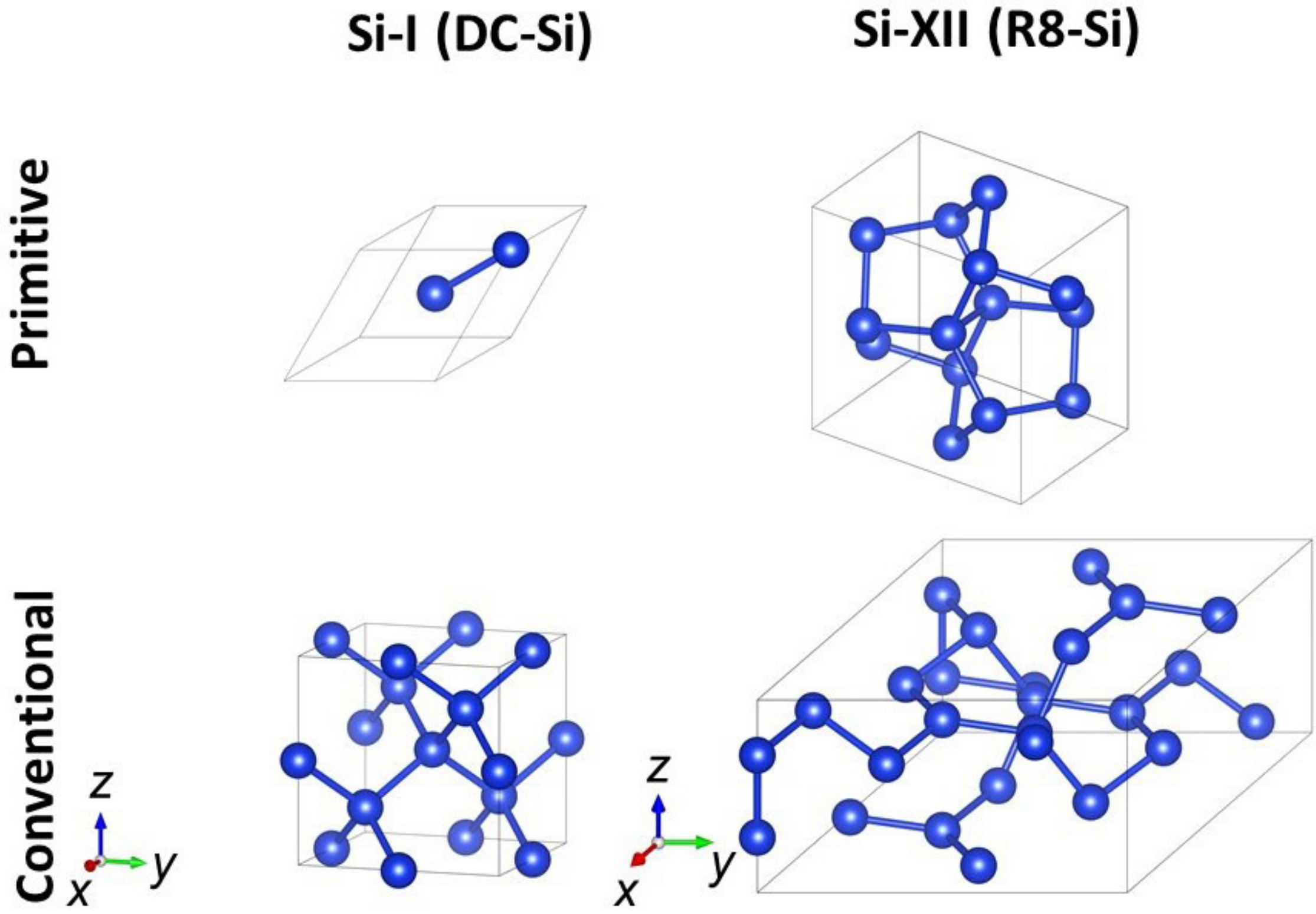}
  \caption{Primitive and conventional unit cell structure of Si$-$I and Si$-$XII. Silicon allotropes stabilized at ambient conditions including DC$-$Si and the most common metastable phases R8$-$Si. }
\end{figure}

The DFT calculations based on projector augmented$-$wave (PAW) \cite{20} were implemented in VASP package \cite{21}. The Perdew$-$Burke$-$Ernzerhof (PBE) form of generalised gradient approximation (GGA) functionals was used for exchange potential. Since the GGA approach underestimates the band gap of a semiconductor, we also performed the Heyd$-$Scuseria$-$Ernzerhof 2006 (HSE06) calculation with the total exchange potential containing 25\% of the Hartree$-$Fock exchange potential to get the accurate band gap and electronic transport properties \cite{22}. The size of plane$-$wave basis set was limited with a cutoff energy of 600~eV. The energy and force convergent criterion were set to be 10$^{-6}$~eV and 10$^{-3}$~eV/\AA~during structure optimization, respectively. Besides, the first Brillouin zone was sampled with $9\times9\times9$ $\Gamma$$-$centred $k$$-$point grid.
For the transport properties calculations, ShengBTE \cite{23} and BoltzTrap2 \cite{24} packages were used to solve the Boltzmann transport equation of phonon and electron, respectively. More computational details can be seen in supplementary material.

\begin{figure}[h]
  \centering
  \includegraphics[width=7cm]{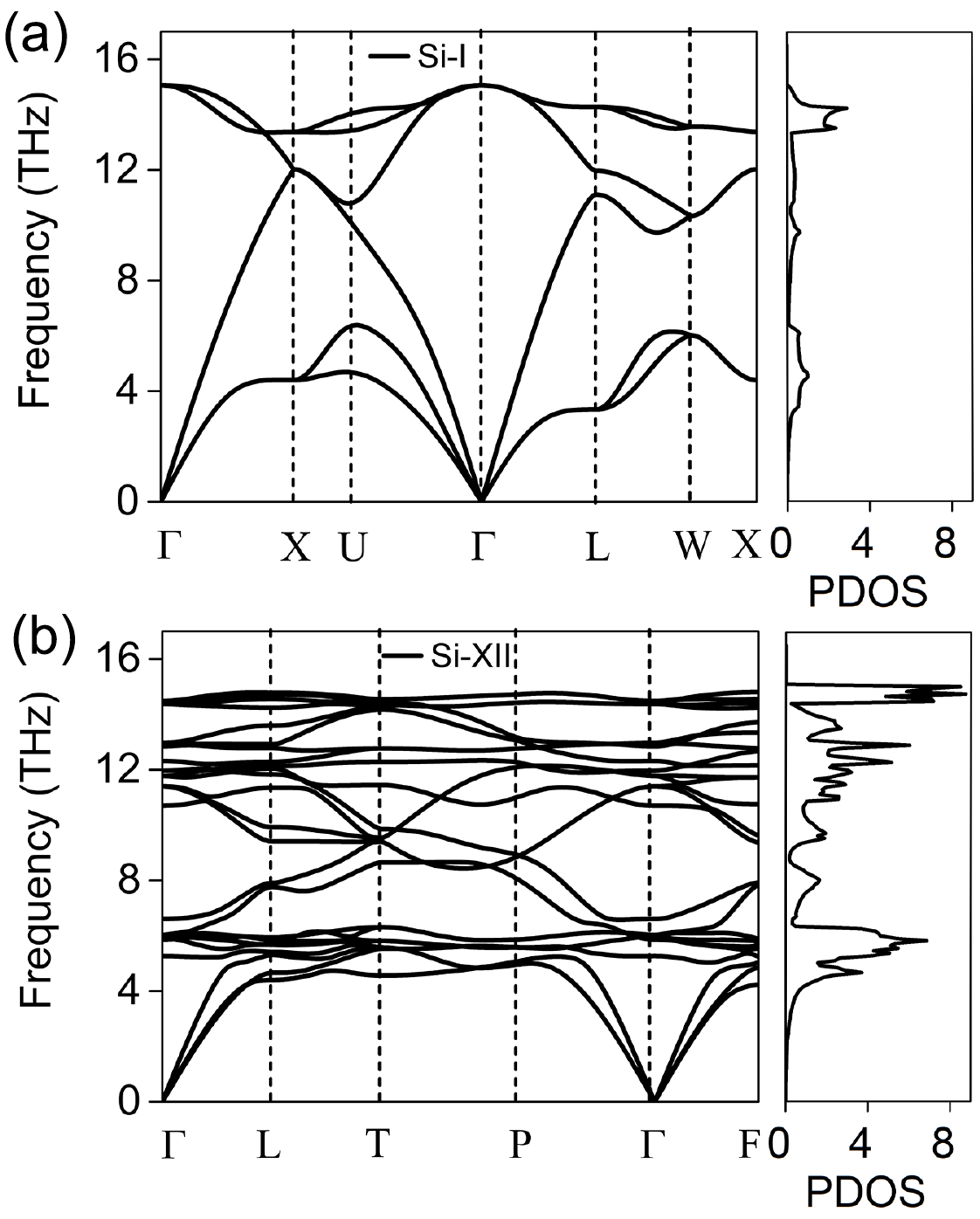}
  \caption{Phonon band structures and phonon density of states of (a) Si$-$I and (b) Si$-$XII, respectively.}
\end{figure}

\begin{figure*}[t]
  \centering
  \includegraphics[width=12cm]{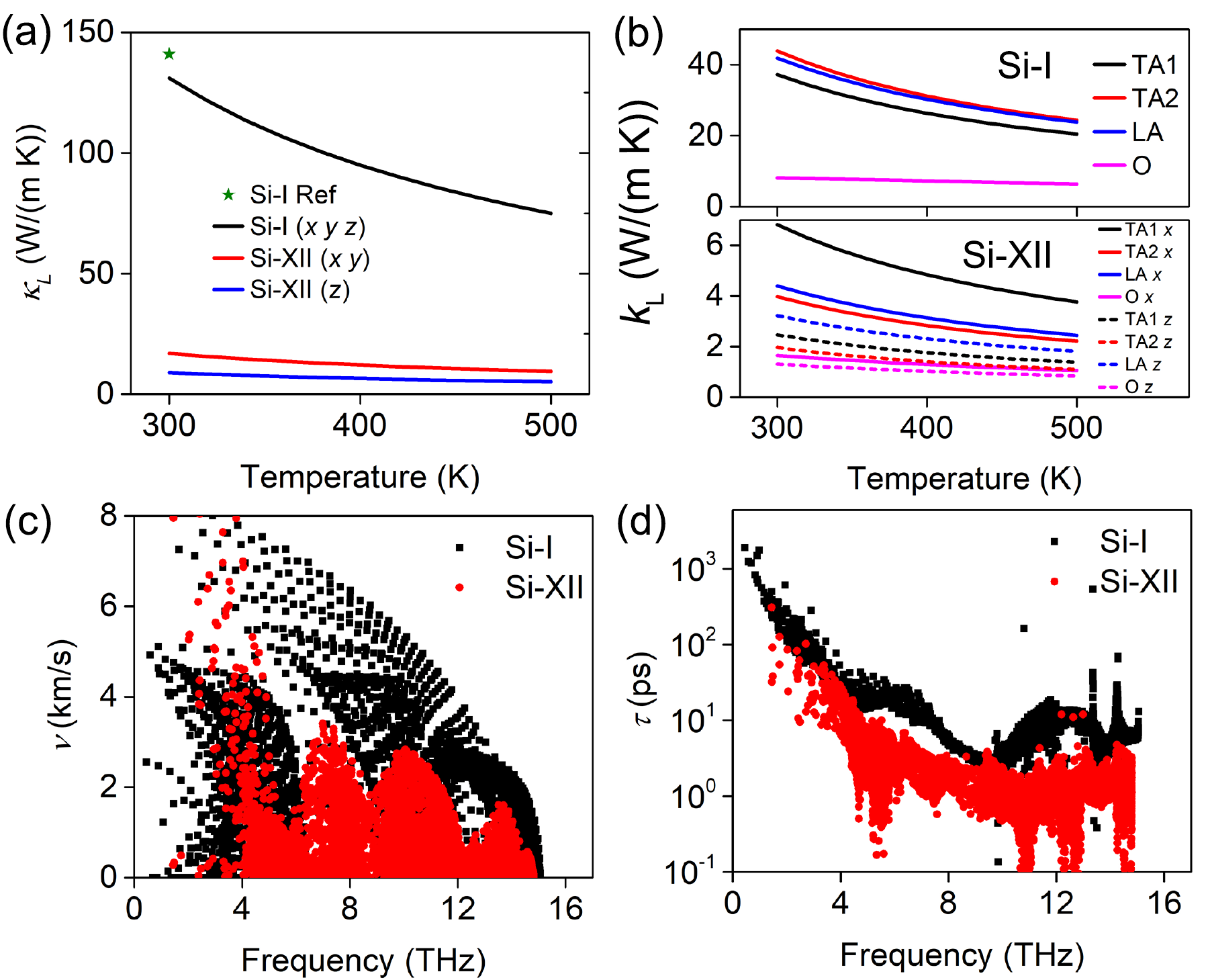}
  \caption{Phonon transport properties in Si$-$I and Si$-$XII. (a) Thermal conductivity, (b) Contribution to lattice thermal conductivity of three acoustic branches and all optical branches, (c) Group velocity, (d) Relaxation time at 300~K.}
\end{figure*}

\begin{figure}
  \centering
  \includegraphics[width=6cm]{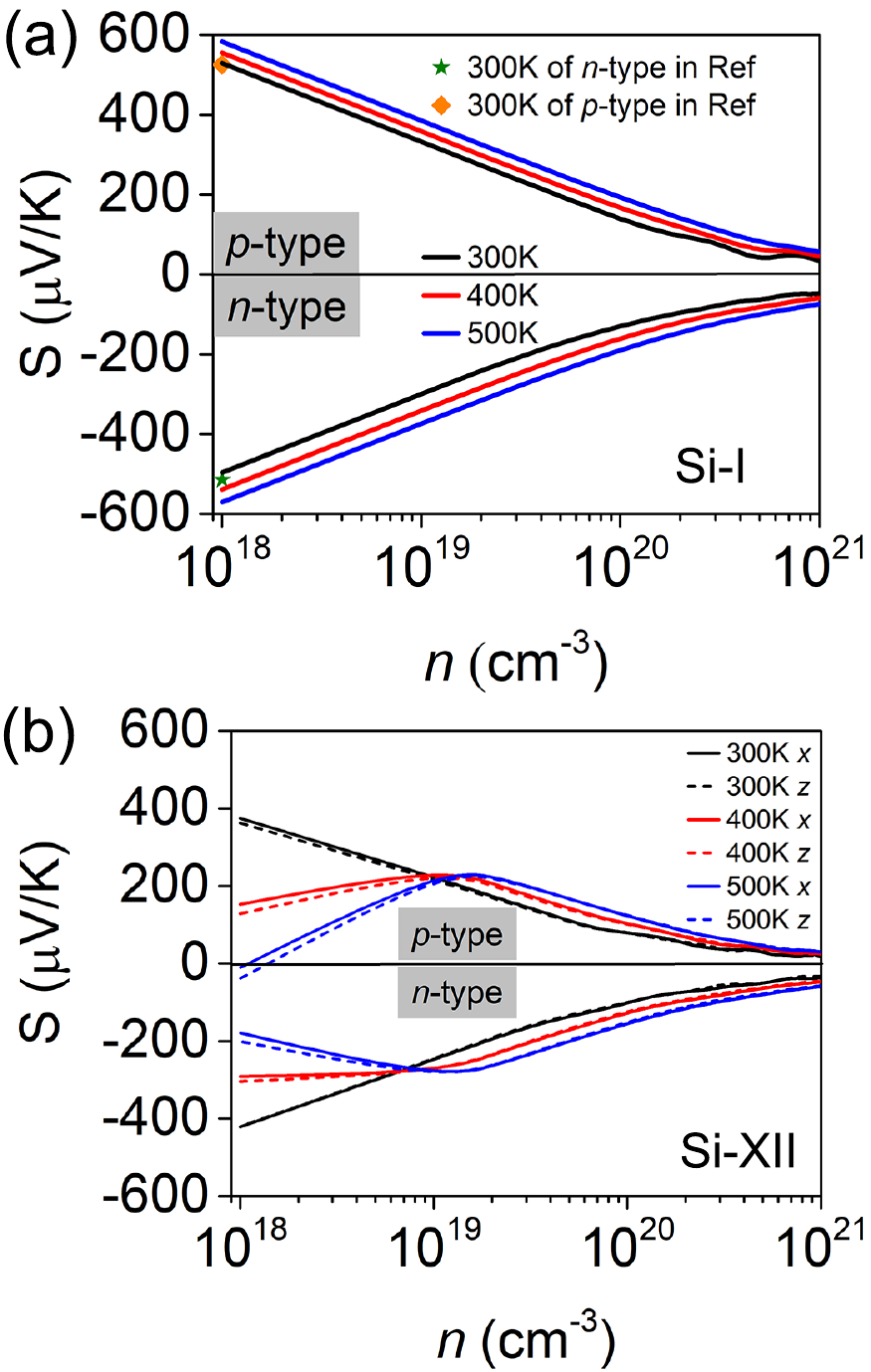}
  \caption{The Seebeck coefficients of (a) Si$-$I and (b) Si$-$XII with $p-$ and $n-$type doping $versus$ carrier concentration at 300, 400, and 500~K, respectively.}
\end{figure}

\begin{figure*}
  \centering
  \includegraphics[width=12cm]{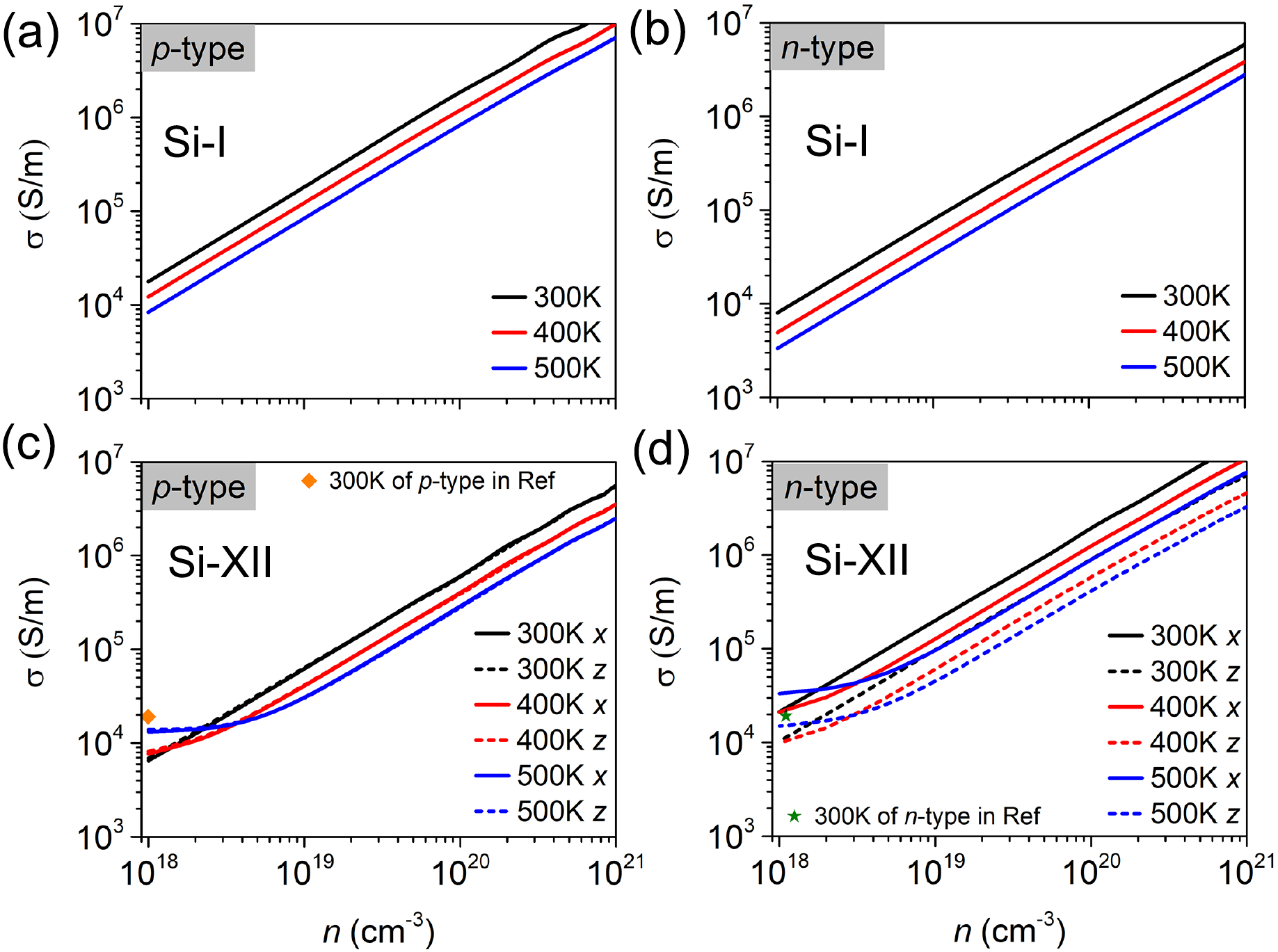}
  \caption{The electrical conductivities of (a) (b) Si$-$I and (c) (d) Si$-$XII with $p-$ and $n-$type doping $versus$ carrier concentration at 300, 400, and 500~K, respectively, in comparison with the calculation results with constant electron relaxation time approximation and PBE functional in Ref. \cite{34}}
\end{figure*}

\begin{figure*}
  \centering
  \includegraphics[width=12cm]{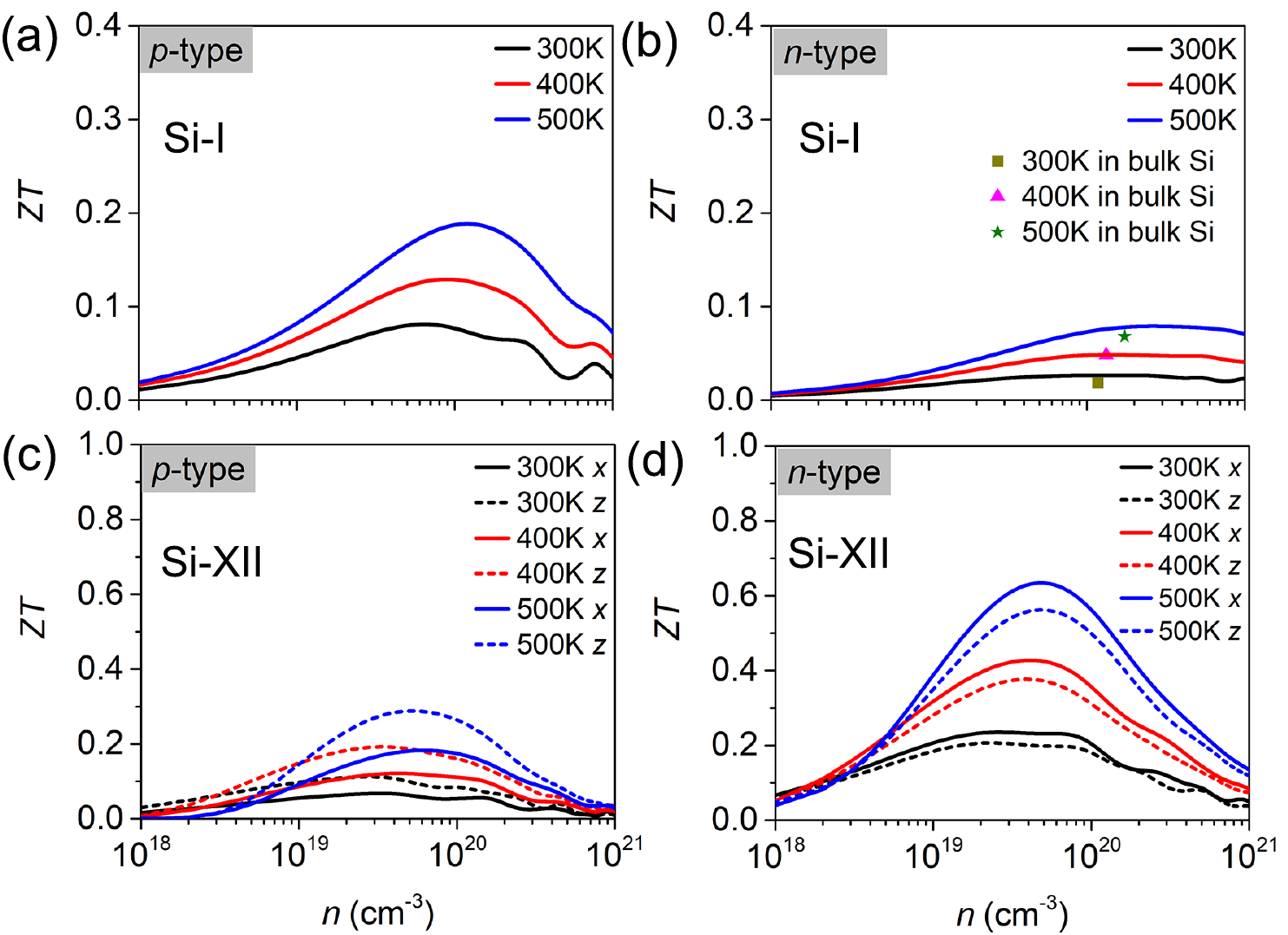}
  \caption{Figure of merit $ZT$ of (a) (b) Si$-$I and (c) (d) Si$-$XII with $p-$ and $n-$type doping $versus$ carrier concentration at 300, 400, and 500~K, respectively.}
\end{figure*}
The optimized lattice constants using PBE functional for primitive Si$-$I and Si$-$XII are 3.87 and 5.77~{\AA}, respectively, in excellent agreement with the reported results \cite{25,26}. Unlike the isotropic Si$-$I, Si$-$XII is an anisotropic system with equivalent $x$ and $y$ directions (see Fig. 1). The phonon dispersion relations of Si$-$I and Si$-$XII are shown in Fig. 2 (a) and (b), respectively. Generally, The phonon branches of Si$-$XII are flatter compared with Si$-$I at high$-$frequency (\textgreater~4~THz). The flat low$-$frequency optical modes are coupled with the acoustic modes, resulting in the stronger phonon scattering by acoustic branches and lower group velocity. Additionally, the high$-$frequency optical phonon branches are strongly interacted by themselves (see the phonon density of states in Fig. 2 (b)) and further reduce the $k_L$.

To quantitatively illustrate the phonon transport performance, Fig. 3 (a) shows the temperature$-$dependent $k_L$ along $x$, $y$, and $z$ axis. $k_L$ for both Si$-$I and Si$-$XII decrease with increasing temperature due to the gradually enhanced phonon scattering strength. It is worth noting that $k_L$ of Si$-$XII is one magnitude smaller than that of Si$-$I. The values for Si$-$I and Si$-$XII are 131.07~Wm${^{-1}}$K${^{-1}}$ (146~Wm${^{-1}}$K${^{-1}}$ in Ref. \cite{27}) and 16.83~Wm${^{-1}}$K${^{-1}}$ along $x$ axis (8.95~Wm${^{-1}}$K${^{-1}}$ along $z$ axis) at 300~K, respectively. Additionally,  the large ratio $k_x$/$k_z$ = 1.88 \textgreater~1 in Si$-$XII indicates the anisotropic heat transport performance.
According to the equation ES1, the $k_L$ is directly determined by the heat capacity $C_v$, group velocity $v$, and phonon lifetime $\tau$. For Si$-$XII, the anisotropic $v$ is the main reason why the $k_L$ shows anisotropy. In Fig. 2 (b), the high symmetry points directions $\Gamma$ (0,0,0) $-$ $L$ (0.5,0,0) and $\Gamma$ (0,0,0) $-$ $F$ (0.5,$-$0.5,0) are along crystal $x$/$y$ axis, $L$ (0.5,0,0) $-$ $T$ (0.5, $-$0.5, 0.5) and $P$ (0.247,0.247,0.247) $-$ $T$ (0.5, $-$0.5, 0.5) directions are along crystal $z$ axis. The $k_L$ is proportional to the $v$ which is defined as $\vec{v}=dw/d\vec{q}$, meaning the flat acoustic phonon branches along $L $–$ T$ and $P $–$ T$ directions result in lower $v$ and $k_L$ along $z$ axis.
Additionally, it was reported that superhard bulk materials with large bulk modulus show high thermal conductivity \cite{28, 29}, and the other factors including low atomic mass, strong interatomic bonding, simple crystal structure, and low anharmonicity possibly play a role in determining the thermal conductivity according to Slack’s criteria \cite{30}. The difference of interatomic bonding strength in different axis directions can also be one of the possible reason why the $k_x$ of Si$-$XII is larger than $k_z$. In Tab. S1, the elastic constant ($C$) along $x$ axis is 168.56~GPa (\textgreater 137.28~GPa along $z$ axis), which confirm the stronger Si$-$Si interatomic bonding strength along $x$ axis.
In Fig. 3 (b), we show the $k_L$ contribution from different phonon branches along different axis. The three acoustic phonons in Si$-$I and Si$-$XII contribute more than 85\% to the total $k_L$ from 300 to 500~K, $e.g.$, at room temperature, the 85\% cumulative $k_L$ of Si$-$XII in $x$ axis depends on phonon branches with frequency below 4.24~THz (shown in Fig. S3). Obviously, the lower phonon group velocity and relaxation time lead to the lower $k_L$ in Si$-$XII (shown in Fig. 3 (c) and (d)).

To evaluate the electron transport properties, the electronic band structure obtained by the HSE06 functional is presented in Fig. S4. Si$-$I shows indirect band gap with the value of 1.17~eV, and the VBM occurs at $\Gamma$ and the CBM occurs between $\Gamma-X$ (here, we set that CBM is located at $P$ in Tab. S1). It is noted that Si$-$XII is also an indirect band gap semiconductor with the value of 0.22~eV, and the VBM and CBM are located at $T$ and $L$, respectively, the data are in good agreement with literatures \cite{26,31}. In particular, the band gap value of Si$-$XII is within the range of traditional TE materials (0.2$-$0.5~eV for Bi$_{\rm 2}${Te$_{\rm 3}$}, PbTe, and CoSb$_{\rm 3}$) \cite{32}, which further motivates us to investigate the TE performance of Si$-$XII.

Fig. 4 (a) and (b) show the Seebeck coefficients as a function of carrier concentration for $p-$ and $n-$type Si$-$I and Si$-$XII at 300, 400, and 500~K, respectively. Owing to the larger band gap of Si$-$I, both $p-$ and $n-$type Si$-$I show larger Seebeck values compared with Si$-$XII within the doping range. The Seebeck coefficient of Si$-$I is inversely proportional to the carrier concentration, in agreement with the Mahan$-$Sofo theory \cite{33}, and the absolute values of $p-$ and $n-$type at 10$^{18}$~cm$^{\rm -3}$ are consistent with Ref. \cite{34} at 300~K.
However, for Si$-$XII at 400 and 500~K, the bipolar effect induced Seebeck coefficients firstly increase and then decrease. Indeed, the bipolar effect on thermoelectric performance is notorious in narrow band gap ($E_g$~\textless~0.5~eV) semiconductors and it originates from the electron excitation from valence band to the conduction band at high temperature \cite{35,36,37,38}. The thermal excitation usually does not change the concentration of major carrier too much but increases the minor carrier concentration. Arising from the two opposite signs of holes and electrons, the total Seebeck coefficients will be canceled out each other, unfavorable for the thermoelectric power generation.

In contrast to the Seebeck coefficient dependence on the carrier concentration and temperature, the electrical conductivity ($\sigma$) is positively correlated with the carrier concentration but negatively correlated with the temperature. It is a general knowledge that the lightweight carrier move faster in nanotransport devices. As presented in Fig. 5, $\sigma$ of $p-$type Si$-$I is higher than that of $n-$type due to the smaller effective mass of hole (the effective mass is listed in Tab. S1), while the larger effective mass of hole in Si$-$XII leads to the lower $\sigma$ of $p-$type doping. Similarly, because of the large effective mass of electron along $z$ axis, $\sigma$ along $z$ axis are obviously lower than that along $x$ axis.

Next, we move to the discussion of electronic thermal conductivity ($k_e$). According to the Wiedemann$-$Franz law \cite{39}, $k_e$ = $L{\sigma}T$, where $L$ is the Lorentz constant (its variation with $n$ and $T$ are shown in Fig. S5), which means $k_e$ is proportional to $\sigma$. So $k_e$ in the Fig. S6 show similar concentration$-$dependent trends like $\sigma$ in Fig. 5. Additionally, the magnitude of $k_e$ is much smaller than that of $k_L$ within low carrier concentration.

Based on the above results of phonon and electronic transport properties, the evaluation of the figure of merit $ZT$ can be written as, $ZT=\frac{{S^2}\sigma}{{k_L}+{k_e}}T$, and the calculated $ZT$ of Si$-$I and Si$-$XII for both $p-$ and $n-$type are shown in Fig. 6. Because doping benefits the electrical conductivity while deteriorating the Seebeck coefficient, $ZT$ firstly increases and then decreases with the carrier concentration. Overall, the $ZT$ of Si$-$XII are remarkedly larger than that of Si$-$I mainly arising from the one magnitude lower $k_L$ in metastable phase. Especially for Si$-$XII in Fig. 6 (c) and (d), the optimal $ZT$ of $n-$type at 300, 400, and 500~K are nearly two times larger than that of $p-$type. Besides, $ZT$ of $n-$type Si$-$XII along $z$ axis are smaller than that along $x$ axis, while $ZT$ of $p-$type show opposite trends, which is directly determined by the electrical conductivities and carrier effective mass. Optimistically, the maximum $ZT$ of $n-$type Si$-$XII along $x$ axis ($z$ axis) at 300, 400, and 500~K are 0.24 (0.21) at 2.6$\times$10$^{19}$~cm$^{\rm -3}$, 0.43 (0.38) at 4.1$\times$10$^{19}$~cm$^{\rm -3}$, 0.63 (0.56) at 4.8$\times$10$^{19}$~cm$^{\rm -3}$, respectively. However, the maximum $ZT$ of doping  Si$-$I is still smaller than 0.2, here, the values at different temperature are also comparable with the other calculation results \cite{40}. Moreover, $ZT$ can be further promoted via suppressing $k_L$, such as involving the grain boundary, isotope effect, and vacancy. Our calculations suggest that metastable Si$-$XII is a promising thermoelectric material.

In summary, by combining the first$-$principles calculations and the Boltzmann transport theory, the TE transport performance of metastable Si$-$XII was comprehensively investigated. For phonon transport aspects, strong phonon anharmonic scattering results in the lower lattice thermal conductivities (16.83 and 8.95~Wm${^{-1}}$K${^{-1}}$ along $x$ and $z$ axis, respectively) of anisotropic Si$-$XII. The electron relaxation times were predicted by the deformation potential theory and then the electronic transport properties were calculated by solving electron Boltzmann transport equation. The results indicated that the low lattice thermal conductivities and high electrical conductivities are beneficial for the improvement of $ZT$ for Si$-$XII. By reasonable carrier doping, the maximum $ZT$ of $n-$type Si$-$XII are evaluated to be 0.24, 0.43, and 0.63 along $x$ axis at 300, 400, and 500~K, respectively. Our work confirms that Si$-$XII is a promising thermoelectric material, and Si$-$XII can be considered to offer fine$-$tuning to further enhance the thermoelectric response in the future theoretical and experimental researches.

\section*{SUPPLEMENTARY MATERIAL}
See supplementary material for computational details, thermodynamic stability of metastable phase, cumulative phonon thermal conductivity, and electron transport properties.

\section*{ACKNOWLEDGMENT}
The authors thank the useful discussions with Junichiro Shiomi from the University of Tokyo and Masamichi Kohno from Kyushu University. This work was supported by the Natural Science Foundation of China (Grant No. 52006134), Shanghai Pujiang Program (Grant No. 20PJ1407500), and Shanghai Key Fundamental Research Grant (No. 21JC1403300). The computations in this paper were run on the $\pi$~2.0 cluster supported by the Center for High Performance Computing at Shanghai Jiao Tong University.

\section*{DATA AVAILABILITY}
The data that support the findings of this study are available from the corresponding author upon reasonable request.


\nocite{*}
\section*{REFERENCES}
\begin{thebibliography}{}
\bibitem{1} J. Wei, L. Yang, Z. Ma, P. Song, M. Zhang, J. Ma, F. Yang, and X. Wang, J. Mater. Sci. {\bf 55}, 12642$-$12704 (2020).
\bibitem{2} L. Zhang, X. Shi, Y. Yang, and Z. Chen, Mater. Today {\bf 46}, 62$-$108 (2021).
\bibitem{3} L. Weber, E. Gmelin, Appl. Phys. A {\bf 53}, 136–140 (1991).
\bibitem{4} S. Ju, X. Liang, and X. Xu, J. Appl. Phys. {\bf 110}, 054318 (2011).
\bibitem{5} V. Kessler, D. Gautam, T. Hülser, M. Spree, R. Theissmann, M. Winterer, H. Wiggers, G. Schierning, and R. Schmechel, Adv. Eng. Mater. {\bf 15}, 379 (2013).
\bibitem{6} A. I. Hochbaum, R. Chen, R. D. Delgado, W. Liang, E. C. Garnett, M. Najarian, A. Majumdar, and P. Yang, Nature {\bf 451}, 163$-$167 (2008).
\bibitem{7} A. I. Boukai, Y. Bunimovich, J. Tahir-Kheli, J. K. Yu, W. A. Goddard Iii, and J. R. Heath, Nature, {\bf 451}, 168$-$171 (2008).
\bibitem{8} J. Shiomi, APL Mater. {\bf 4}, 104504 (2016).
\bibitem{9} D. Li, Y. Wu, P. Kim, L. Shi, P. Yang, A. Majumdar, Appl. Phys. Lett. {\bf 83}, 2934$-$2936 (2003).
\bibitem{10} R. Chen, A. I. Hochbaum, P. Murphy, J. Moore, P. Yang, A. Majumdar, Phys. Rev. Lett. {\bf 101}, 105501 (2008).
\bibitem{11} B. Haberl, M. Guthrie, S. V. Sinogeikin, G. Shen, J. S. Williams, and J. E. Bradby, High Pressure Res. {\bf 35}, 99 (2015).
\bibitem{12} B. Haberl, T. A. Strobel, and J. E. Bradby, Appl. Phys. Rev. {\bf 3}, 040808 (2016).
\bibitem{13} Z. Zeng, Q. Zeng, W. L. Mao, and S. Qu, J. Appl. Phys. {\bf 115}, 103514 (2014).
\bibitem{14} J. C. Jamieson, Science {\bf 139}, 762 (1963).
\bibitem{15} R. H. Wentorf and J. S. Kasper, Science {\bf 139}, 338 (1963).
\bibitem{16} B. Chon, Y. Ikoma, M. Kohno, J. Shiomi, M. R. McCartney, D. J. Smith, and Z. Horita, Scr. Mater. {\bf 157}, 120 (2018).
\bibitem{17} S. Wippermann, M. Vörös, D. Rocca, A. Gali, G. Zimanyi, and G. Galli, Phys. Rev. Lett. {\bf 110}, 046804 (2013).
\bibitem{18} H. Zhang, H. Liu, K. Wei, O. O. Kurakevych, Y. Le Godec, Z. Liu, J. Martin, M. Guerrette, G. S. Nolas, and T. A. Strobel, Phys. Rev. Lett. {\bf 118}, 146601 (2017).
\bibitem{19} C. Shao, K. Matsuda, S. Ju, Y. Ikoma, M. Kohno, and J. Shiomi, J. Appl. Phys. {\bf 129}, 085101 (2021).
\bibitem{20} P. E. Blöchl, Phys. Rev. B {\bf 50}, 17953 (1994).
\bibitem{21} G. Kresse and J. Furthmüller, Phys. Rev. B {\bf 54}, 11169 (1996).
\bibitem{22} J. Heyd, G. E. Scuseria, and M. Ernzerhof, J. Chem. Phys. {\bf 118}, 8207 (2003).
\bibitem{23} W. Li, J. Carrete, N. A. Katcho and N. Mingo, Comput. Phys. Commun. {\bf 185}, 1747 (2014).
\bibitem{24} G. K. H. Madsen, J. Carrete, and M. J. Verstraete, Comput. Phys. Commun. {\bf 231}, 140 (2018).
\bibitem{25} L. He, F. Liu, G. Hautier, M. J. T. Oliveira, M. A. L. Marques, F. D. Vila, J. J. Rehr, G.-M. Rignanese, and A. Zhou, Phys. Rev. B {\bf 89}, 064305 (2014).
\bibitem{26} B. G. Pfrommer, M. Cote, S. G. Louie, and M. L. Cohen, Phys. Rev. B {\bf 56}, 6662 (1997).
\bibitem{27} B. Liao, B. Qiu, J. Zhou, S. Huberman, K. Esfarjani, and G. Chen, Phys. Rev. Lett. {\bf 114}, 115901 (2015).
\bibitem{28} J. Liu, S. Ju, N. Nishiyama, and J. Shiomi, Phys. Rev. B {\bf 100}, 064303 (2019).
\bibitem{29} S. Ju, R. Yoshida, C. Liu, S. Wu, K. Hongo, T. Tadano, and J. Shiomi, Phys. Rev. Mater. {\bf 5}, 053801 (2021).
\bibitem{30} G. A. Slack, J. Phys. Chem. Solids {\bf 34}, 321 (1973).
\bibitem{31} M. S. Hybertsen and S. G. Louie, Phys. Rev. B {\bf 34}, 5390 (1986).
\bibitem{32} J. Shen, Z. Chen, L. Zheng, W. Li, Y. Pei, J. Mater. Chem. C {\bf 4}, 209 (2016).
\bibitem{33} G. J. Snyder, E. S. Toberer, Nat. Mater. {\bf 7}, 105 (2008).
\bibitem{34} F. Ricci, W. Chen, U. Aydemir, G. Jeffrey Snyder, G. Rignanese, A. Jain, and G. Hautier, Sci. Data {\bf 4}, 170085 (2017).
\bibitem{35} J. H. Bahk and A. Shakouri, Appl. Phys. Lett. {\bf 105}, 052106 (2014).
\bibitem{36} L. D. Zhao, H. J. Wu, S. Q. Hao, C. I. Wu, X. Y. Zhou, K. Biswas, J. Q. He, T. P. Hogan, C. Uher, C. Wolverton, V. P. Dravidc, and M. G. Kanatzidis, Energy. Environ. Sci. {\bf 6}, 3346 (2013).
\bibitem{37} J. J. Gong, A. J. Hong, J. Shuai, L. Li, Z. B. Yan, Z. F. Renb, and J.-M. Liu, Phys. Chem. Chem. Phys. {\bf 18}, 16566 (2016).
\bibitem{38} J. H. Bahk and A. Shakouri, Phys. Rev. B {\bf 93}, 165209 (2016).
\bibitem{39} M. Jonson and G. Mahan, Phys. Rev. B {\bf 21}, 4223 (1980).
\bibitem{40} G. H. Zhu, H. Lee, Y. C. Lan, X.W. Wang, G. Joshi, D. Z. Wang, J. Yang, D. Vashaee, H. Guilbert, A. Pillitteri, M. S. Dresselhaus, G. Chen, and Z. F. Ren, Phys. Rev. Lett. {\bf 102}, 196803 (2009).
\end{thebibliography}
\end{document}